# Novel aspects of higher dimensional structural description of trigonal, pentagonal and their related phases

R.K. Mandal

Centre of Advanced Study
Department of Metallurgical Engineering
Institute of Technology, Banaras Hindu University
Varanasi-221005 (India)
Email : rkmandal21@yahoo.com; rkmandal.met@itbhu.ac.in

**Abstract**

A four dimensional description for the trigonal phase is presented. It has been demonstrated that it is possible to model one dimensional quasiperiodic structure with trigonal symmetry apart from recovering the Miller-Bravais scheme for the description of hexagonal phases. A similar discussion on the two classes of decagonal phases having 10/m and $10_5$ symmetries will be given. The distinction in the six dimensional models of the two classes of solids has been pointed out. The zone rules for all the cases have been formulated. The critical comparison for the structural description of trigonal, decagonal and related phases in terms of higher dimension is made. The unification achieved in the higher dimensional structural models of various phases has been emphasized. The importance of such models for the study of structural phase transformation has been indicated.

## 1. Introduction

Symmetry apart from imparting sense of aesthetics to natural organization provides easier way of visualizing underlying atomic arrangements of solids. The point group and space group symmetries help attain mathematically exact description of atomic arrangements of solids having long range translational orders. They are characterized by the existence of sharp diffraction peaks in reciprocal space. All of them have corresponding diffraction vectors. They are measured with respect to a chosen origin located at the centre of transmitted beam. All the diffraction vectors or reciprocal lattice vectors are expressed as integral linear combination of basis vectors. The minimum number of integrally independent vectors is known as rank of the solids (Mermin 1992a , 1992b). If one collects the integral coefficients of a diffraction vector, then every peak is represented by a unique set of indices. The minimum number of indices utilized for the purpose is equal to the rank of the solids. All three dimensional periodic solids have rank equal to three. A triplet of indices (hkl), therefore, corresponds to a diffraction peak. For hexagonal crystal, if same viewpoint is adopted, one encounters a problem pertaining to symmetry. This relates to the fact that diffraction vectors related by six fold symmetry possess a set of triplets whose indices are not obtainable through their permutation. This is avoided by introducing four indices Miller-Bravais (MB) scheme. Frank (1965) and Mackay (1977) have discussed various aspects of this indexing system. Frank (1965) invoked the concept of higher dimensional space to rationalize the underlying philosophy of utilizing four basis vectors in direct and reciprocal spaces. In contrast to this, Mackay (1977) utilized the method of Moore-Penrose generalized inverse to illustrate the relationship between direct and reciprocal bases in Miller-Bravais scheme. This technique is helpful for incommensurate and quasiperiodic structures (Mandal 1994, Lord 2003).

The discovery of decagonal phases (Chattopadhyay et al. 1985, Bendersky 1985) also posed a problem akin to that of hexagonal crystals. Decagonal phases are quasiperiodic in two dimensions (d) and periodic in a direction perpendicular to the quasiperiodic plane. The rank of such a solid is five. Four basis vectors oriented with respect to each other by $72^0$ and the fifth one perpendicular to this plane are sufficient to map the entire reciprocal space. However, a set of vectors related by five fold symmetry may not possess quintuplet of indices that are permuted. Ranganathan et al. (2007) have deliberated extensively on the importance of having M-B scheme for decagonal phases.

The additional vector required to preserve symmetry in the indices for a family of directions/planes during indexing gives rise to the problem of redundancy. This refers to the non-unique assignment of indices to a diffraction spot. This can be surmounted by putting condition on the permissible set of indices. The M-B scheme demands that sum of the indices corresponding to planar vector be kept equals to zero. It will be shown that such a choice is too restrictive. The general condition may permit us to overcome the problem of non-uniqueness and also help us interpret newer phases. Such a view point is lacking in literature. This will be substantiated by revisiting the MB indexing schemes of hexagonal (Frank, 1965; Mackay, 1977), decagonal (Ranganathan et al. 2007) and their related phases with the help of higher dimensional approach. The 1d aperiodic phase having trigonal symmetry will also be discussed. Lifshitz and Mermin (1994a, 1994b) have extensively deliberated on the Bravais classes and space groups of trigonal and hexagonal quasiperiodic phases. Their approach is based on the Copernican crystallography developed in Fourier space ( Bienenstock and Ewald 1962; Mermin 1992). In contrast, the canonical cut and project scheme ( de Wolff et al. 1981; Baake and Moody 2004) is capable of providing information about the atomic positions as well as the intensity of quasiperiodic structures (Henley et al.2006). Our purpose will be to bring out the effects of metrical and symmetrical properties of the 4d crystal on the structures that arise in parallel or physical space with the help this approach. This presentation will also bring out essential distinction between two classes of 2d decagonal phases having 10/m and $10_5$ symmetries in terms of six dimensional models. The limiting cases of 4d and 6d structural description for the trigonal and decagonal phases respectively seem to offer striking similarities in view of parallel and perpendicular space characteristics. Such aspects have not been discussed in relation to canonical cut and project method for modeling of experimentally observed quasicrystalline phases.

## 2. Four dimensional structural models for trigonal and related phases

In this presentation, a phase will be said to be related to each other if group-subgroup relationships between the parent and product phase exist. A solid with trigonal symmetry is obviously related to a solid having hexagonal symmetry. Similarly, there may be other possibilities that will be discussed later in this article. Such structures can be described on a unified basis if a general framework is considered. For achieving this, the following set of basis vectors may be taken.

$$A_i^{\parallel} = |A_1^{\parallel}| \, [\sin\theta_h \, T^{i-1} X^{\parallel} + \cos\theta_h \, Z^{\parallel}]$$

$$A_4^{\parallel} = |A_4^{\parallel}| \, Z^{\parallel} \qquad (1)$$

Where i = 1, 2, 3 and T stands for rotation by 120° in anticlockwise direction.

$X^{\parallel}$, $Y^{\parallel}$ and $Z^{\parallel}$ are Cartesian bases in physical space. The use of parallel (||) has been made to indicate that all the vectors are in the physical space or parallel space. This terminology is quite familiar in the higher dimensional structural description of quasicrystals (Elser and Henley, 1985; Duneau and Katz, 1985). The value of $\theta_h$ is the angle between the Z″ axis (parallel to trigonal axis) and $A_i^{\parallel}$

(i = 1 to 3). For $\theta_h = 90°$, $A_i^{\|}$ (i = 1 to 3) are parallel to $A_i$ (i = 1 to 3) of Fig. 1a. It may be noted that $\theta_h$ provides a parameter to model structures of many related trigonal phases.

Following the algorithm of constructing projection matrix proposed by Lele and Mandal 1992 for the vacancy ordered τ phases, one may demonstrate that the magnitude of basis vectors are:

$$|A_1^{\|}| = \sqrt{\frac{2}{3\sin^2\theta_h}} \;;$$

$$|A_4^{\|}| = \sqrt{\frac{1-3\cos^2\theta_h}{\sin^2\theta_h}} \qquad (2)$$

It is obvious that $\cos^{-1}1/\sqrt{3} \geq \theta_h \leq 90°$. For $\theta = 1/\sqrt{3}$, $|A_4^{\|}| = 0$ and fourth vector vanishes. Below this, the formalism breaks down. As stated earlier at $\theta_h = 90°$; the bases are analogous to those originally proposed in MB scheme (Frank, 1965; Mackay, 1977). The projection matrix P of $\|$ space is given by

$$P = \begin{bmatrix} P_{11} & P_{12} & P_{13} & P_{14} \\ P_{12} & P_{11} & P_{12} & P_{14} \\ P_{13} & P_{12} & P_{11} & P_{14} \\ P_{14} & P_{14} & P_{14} & P_{14} \end{bmatrix} \qquad (3)$$

Where $P_{11} = |A_1^{\|}|^2$; $P_{12} = P_{13} = |A_1^{\|}|^2 [-\frac{\sin^2\theta_h}{2} + \cos^2\theta_h]$;

$P_{14} = |A_1^{\|}||A_4^{\|}| \cos\theta_h$; $P_{44} = |A_4^{\|}|^2$.

The corresponding projection matrix Q in complementary or perpendicular or pseudo space (⊥) is obtained by Q = I–P where I is an (4×4) identity matrix. This gives rise to ⊥ space bases as

$$A_1^{\perp} = A_2^{\perp} = A_3^{\perp} = |A_1^{\perp}| Z^{\perp}$$

$$A_4^{\perp} = -|A_4^{\perp}| Z^{\perp} \qquad (4)$$

Where $|A_1^{\perp}| = \sqrt{\dfrac{1-3\cos^2\theta_h}{3\sin^2\theta_h}}$ ; $|A_4^{\perp}| = \cot\theta_h \sqrt{2}$ .

Bases given in set of equations (1) and (4) respectively for || and ⊥ spaces help us construct an orthogonal hyperlattice in 4 dimensional with hyperlattice parameters 'a' and 'c'. If $e_i$ (i = 1 to 4) are orthonormal basis vectors for 4d orthogonal hyperlattice then the 4d reciprocal lattice vector is given by

$$G^4 = \frac{1}{a}\sum_1^3 n_i e_i + \frac{1}{c} n_4 e_4 \qquad (5)$$

Where $n_1$, $n_2$, $n_3$ and $n_4$ are indices of reflection.

The || and ⊥ components of $G^4$ are respectively designated as $G^{\parallel}$ and $G^{\perp}$ . They follow from equations (1) and (4) as

$$G^{\parallel} = \frac{1}{a}\sum_1^3 n_i A_4^{\parallel} + \frac{1}{c} n_4 A_4^{\parallel} \qquad (6)$$

$$G^{\perp} = \frac{1}{a}\sum_1^3 n_i A_i^{\perp} + \frac{1}{c} n_4 A_4^{\perp} \qquad (7)$$

In an analogous way, the 4d direct lattice vector is of the form

$$R^4 = a \sum_{1}^{3} m_i e_i + m_4 e_4 \quad (8)$$

Where, $m_1$, $m_2$, $m_3$ and $m_4$ are integers. The || and ⊥ components of $R^4$ are given by

$$R^{\parallel} = a \sum_{1}^{3} m_i A_i^{\parallel} + c\, m_4 A_4^{\parallel} \quad (9)$$

$$R^{\perp} = a \sum_{1}^{3} m_i A_i^{\perp} + c\, m_4 A_4^{\perp} \quad (10)$$

Having formulated a general frame work, it is important to recover all the attributes of MB scheme as given in Frank, 1965. The zone rule in || space for $\theta_h = 90°$ is a good starting point for this. The product of

$$G^{\parallel}R^{\parallel} = P_{11} \sum_{i=1}^{3} n_i m_i + P_{44} n_4 m_4 + P_{12}[n_1 m_2 + n_1 m_3 + n_2 m_1 + n_2 m_3 + n_3 m_1 + n_3 m_2] +$$

$$P_{14}\left[(c/a)\, m_4 \sum_{i}^{3} n_i + (a/c)\, n_4 \sum_{1}^{3} m_i\right] \quad (11)$$

For $\theta_h = 90°$; $P_{11} = |A_1^{\parallel}|^2 = \frac{2}{3}$; $P_{12} = -|A_1^{\parallel}|^2/2$; $P_{14} = 0$ and $P_{44} = 1$. Therefore,

$$G^{\parallel} \cdot R^{\parallel} = \frac{2}{3} \sum_{1}^{3} n_i m_i + n_4 m_4 - \frac{1}{3}[n_1(m_2 + m_3) + n_2(m_1 + m_3) + n_3(m_1 + m_2)] \quad (11a)$$

Frank, 1965 has shown based on 4d formalism the elegance that one gets by imposing

$\sum_{1}^{3} n_i = 0$ and $\sum_{1}^{3} m_i = 0$. These substitutions lead to

$$G^{\|} \cdot R^{\|} = \sum_{1}^{4} n_i m_i \qquad (11b)$$

A set of reflections $\{n_1\, n_2\, n_3\, n_4\}$ will be lying in the zone $< m_1\, m_2\, m_3\, m_4 >$ if

$$\sum_{1}^{4} n_i m_i = 0 \qquad (11c)$$

This is the zone axis for the MB scheme.

Please note that for $\theta_h = 90°$; $|A_4^{\perp}| = 0$ and ensuring $\sum_{1}^{3} n_i = 0$; $\sum_{1}^{3} m_i = 0$ leads to retention of 3d physical space only. This is a special 3d-section of the 4d crystal (Frank, 1965). It is important to remember that this is a special choice even for $\theta_h = 90°$ (cf. Eq. 10a). The basis vectors given in equation (1) for $\theta_h \neq 90°$ have the following relation

$$\sum_{1}^{3} A^{\|} = 3 \cos\theta_h \, |A_1^{\|}| \, Z^{\|} \qquad (11d)$$

The right hand side of Equation (11d) goes to zero for $\theta_h = 90°$ and $A_i^{\|}$ (i = 1 to 3) are parallel to planar bases as shown in Figure (1b). For all those acceptable values of $\theta \neq 90°$, it is non-zero, in the plane and has a component parallel to the trigonal axis (or $Z^{\|}$-axis). Hence $\sum_{1}^{3} n_i$ and $\sum_{1}^{3} m_i$ cannot be set to zero. The $Z^{\|}$ component of $G^{\|}$ is

$$G_z^{\parallel} = \frac{|A_1^{\parallel}|\cos\theta_h}{aq}\left[q\sum n_i + p\,n_4\right] \qquad (12)$$

Where $\frac{a}{c}\,\frac{|A_4^{\parallel}|}{|A_1^{\parallel}|\cos\theta_h} = \frac{p}{q}$. A typical case may be taken to demonstrate the generality of the present approach for $\cos\theta_h = 1/2$; p/q = 2 with $\frac{a}{c} = 2\sqrt{2/3}$.

Hence $$G_z^{\parallel} = \frac{|A_1^{\parallel}|\cos\theta_h}{a}\left[\sum n_i + 2n_4\right] \qquad (13)$$

It is seen that

$$G^{\parallel}(222\bar{3}) = 0 \qquad (14)$$

For unique indexing of the trigonal phase,

$$\sum_1^4 n_i = 0, \pm 1 \qquad (15)$$

The Z-component of $R^{\parallel}$ for this case is

$$R_z^{\parallel} = \frac{a}{18|A_1^{\parallel}|\cos\theta_h}\left[4\sum m_i + 3m_4\right] \qquad (16)$$

This leads to

$$R^{\parallel}(111\bar{4}) = 0 \qquad (17)$$

This imposes condition of $m_i$'s for unique representation of a direct space vector as

$$\sum_{1}^{4} m_i = 0 \qquad (18)$$

The zone rule for this case is given by

$$20\sum_{1}^{3} n_i m_i - 2\sum_{1}^{4} n_i m_i + 5 m_4 \sum_{1}^{3} n_i = 0 \qquad (19)$$

At the limiting point of $\cos\theta_h = \frac{1}{\sqrt{3}}$; $|A_1^{\parallel}| = |A_2^{\parallel}| = |A_3^{\parallel}| = 1$; $|A_4^{\parallel}| = 0$; $P_{12} = P_{13} = 0$. Hence the bases are orthonormal. This means the cubic symmetry is recovered. In $\perp$ space, this leads to $|A_1^{\perp}| = 0$; $|A_4^{\perp}| = 1$.

The 4d hyperlattice parameter c, therefore contributes to $\perp$ space only. Thus, the bases have the capability to interpret all those structures that are related to the geometrical and symmetrical properties of cubic and hexagonal phases. Such a unified model may be helpful in understanding the structural transformations possible in intermetallic phases. However, this aspect will not be explored in this communication.

It may be pointed out here that there are many other choices of cos θ for which $G_z^{\parallel}$ and $R_z^{\parallel}$ can be aperiodic. The value of a/c can be assumed to be same as $\sqrt{8/3}$. As a consequence of this, a trigonal structure with aperiodicity along the symmetry axis can be modeled. One of the choices could be $\cos\theta_h = \frac{1}{\sqrt{5}}$. This value is closer to $\cos\theta_h = 1/2$. It has been explained by Mandal and Lele 1989; 1991 for decagonal phases that 2 is the first rational approximant of $\sqrt{5}$.

## 3. Six dimensional structural description of pentagonal and decagonal phases

It has been emphasized earlier by us (Mandal and Lele 1989; 1991 and Mandal et al. 2004) that a distorted icosahedral basis vectors are the best to establish relation between the icosahedral and two dimensionally quasiperiodic structures having 10-fold and 5-fold symmetries. A distortion along one of the five axes of an icosahedron preserves a five-fold symmetry along it. For continuity, the basis vectors utilized by us earlier are reproduced below:

$$V_i = |V_1| [\sin\theta_D T^{i-1} X^{\parallel} + \cos\theta_D Z^{\parallel}]$$

$$V_6 = |V_6| Z^{\parallel} \qquad (20)$$

where $V_i$ (i = 1 to 5) are parallel to the vertex vector of an icosahedron and $V_6$ is the sixth vertex vector along which distortion can be given by taking value of cos $\theta_D$ different from $\frac{1}{\sqrt{5}}$. For this choice, an ideal icosahedral basis vectors are recovered in physical space (Elser & Henley 1985; Duneau & Katz, 1985). A symmetric projection matrix can be constructed through the dot product of vectors given in equation (20). The matrix P has the following form

$$P = \begin{bmatrix} P_{11} & P_{12} & P_{13} & P_{13} & P_{12} & P_{16} \\ P_{12} & P_{11} & P_{12} & P_{13} & P_{13} & P_{16} \\ P_{13} & P_{12} & P_{11} & P_{12} & P_{13} & P_{16} \\ P_{13} & P_{13} & P_{12} & P_{11} & P_{12} & P_{16} \\ P_{12} & P_{13} & P_{13} & P_{12} & P_{11} & P_{16} \\ P_{16} & P_{16} & P_{16} & P_{16} & P_{16} & P_{66} \end{bmatrix} \qquad (21)$$

where $P_{11} = V_1 \cdot V_1 = \dfrac{2}{5\sin^2\theta_D}$; $\quad P_{12} = P_{11}\left[\dfrac{1}{2\tau}\sin^2\theta_D + \cos^2\theta_D\right]$;

$P_{13} = P_{11}\left[-\dfrac{\tau}{2}\sin^2\theta_D + \cos^2\theta_D\right]$; $\quad P_{16} = |V_1||V_1|\cos\theta_D$;

$P_{44} = V_6 \cdot V_6 = \dfrac{5}{2}|V_1|^2(1 - 3\cos^2\theta_D)$ and $\tau = \dfrac{\sqrt{5}+1}{2}$.

This matrix permits 6d orthonormal basis to define the physical or parallel (||) space bases as given in equation (20). For details, readers are referred to our earlier work (Mandal and Lele 1989). The projection matrix Q in the complementary or pseudo or perpendicular (⊥) space is given by Q = I–P where I is an identity matrix of order 6. The corresponding basis vectors in ⊥ space are written as

$$W_i = |W_1|[\sin\phi \; T^{\langle 2i-1\rangle} X^{\perp} + \cos\phi \; Z^{\perp}]$$

$$W_6 = |W_6| \; Z^{\perp}$$

where $|W_1|^2 = 1 - |V_1|^2$; $|W_6|^2 = 1 - |V_6|^2$; $\cos\phi = \left[\frac{1 - 3\cos^2\theta_D}{3 - 5\cos^2\theta_D}\right]^{1/2}$ and i = 1 to 5 and $\langle 2i - 2\rangle$ is modulo 5. The 6d reciprocal lattice vector $G^6$ in terms of orthonormal basis vectors ($e_i$ for i = 1 to 6) for orthogonal cell (Mandal and Lele, 1989) is written as

$$G^6 = \frac{1}{t_1}\sum_1^5 N_i e_i + \frac{1}{t_6} N_6 e_6 \qquad (23)$$

where $t_1$ and $t_6$ are hyperlattice parameters for the 6d cell.

The 3d parallel (||) and perpendicular (⊥) spaces are denoted here by $G^{\|}$ and $G^{\perp}$ respectively. They are given by following equations

$$G^{\|} = \frac{1}{t_1}\sum_1^5 N_i V_i + \frac{1}{t_6} N_6 V_6 \qquad (24)$$

$$G^{\perp} = \frac{1}{t_1}\sum_1^5 N_i W_i + \frac{1}{t_6} N_6 W_6 \qquad (25)$$

where $N_i$'s are indices of reflections.

Similarly, the 6d direct space lattice vector ($R^6$) is written as

$$R^6 = t_1 \sum_1^5 M_i e_i + t_6 M_6 e_6 \qquad (26)$$

The parallel and perpendicular space components of $R^6$ are designated here by $R^{\parallel}$ and $R^{\perp}$ respectively. They are depicted in the form of equations below.

$$R^{\parallel} = t_1 \sum_{1}^{5} M_i V_i + t_6 M_6 V_6 \tag{27}$$

$$R^{\perp} = t_1 \sum_{1}^{5} M_i W_i + t_6 M_6 W_6 \tag{28}$$

where $M_i$ (i = 1 to 6) are integers. The product

$$G^{\parallel} \cdot R^{\parallel} = P_{11} \sum_{1}^{5} M_i N_i + P_{66} M_6 N_6 + P_{12} N_{12} + P_{13} N_{13}$$

$$+ \frac{t_1}{t_6} P_{16} N_6 \sum_{1}^{5} M_i + \frac{t_6}{t_1} P_{16} M_6 \sum_{1}^{5} N_i \tag{29}$$

where
$$N_{12} = N_1(M_2 + M_5) + N_2(M_1 + M_3) + N_3(M_2 + M_4) + N_4(M_3 + M_5) + N_5(M_1 + M_4)$$

$$N_{13} = N_1(M_3 + M_4) + N_2(M_4 + M_5) + N_3(M_5 + M_1) + N_4(M_1 + M_2) + N_5(M_2 + M_3)$$

For the icosahedral phase $\cos\theta_D = \frac{1}{\sqrt{5}}$ and $t_1 = t_6$, hence $P_{11} = \frac{1}{2}$; $P_{12} = \frac{1}{2\sqrt{5}}$; $P_{13} = -\frac{1}{2\sqrt{5}}$; $P_{16} = \frac{1}{2\sqrt{5}}$ and $P_{66} = \frac{1}{2}$. This reduces equation (29) as

$$2\sqrt{5}\, G^{\|} \cdot R^{\|} = \sqrt{5} \sum_{1}^{6} M_i N_i + (N_{12} - N_{13} + N_{16}) \qquad (30)$$

where $$N_{16} = N_6 \left[ \sum_{1}^{6} M_i + M_6 \sum_{1}^{5} N_i \right]$$

The right hand side of equation (30) has rational and irrational parts. Hence, left hand side cannot be equated to zero to recover exact zone rule that is applicable for crystals, (cf. equation 19). However, this can be made to accept values nearer to zero and for special set of $(N_1 N_2 N_3 N_4 N_5 N_6)$ corresponding to chosen symmetric direction (like 2-fold, 3-fold and 5-fold), it may display exactly zero. The purpose of this section is to discuss the structural characteristics of 2d-quasiperiodic structures and aspects pertaining to 3d icosahedral phases can be found in literature (Cahn et al. 1986).

The planar pentagonal scheme (Fitz Gerald et al. 1988, Singh and Ranganathan, 1996a,b; Lord and Mukhopadhyay 2002; Ranganathan et al. 2007), in the model of Mandal and Lele 1989 corresponds to $\theta_D = 90^{\circ}$. This leads to $P_{11} = 2/5$; $P_{12} = \frac{1}{5\tau}$; $P_{13} = -\tau/5$; $P_{16} = 0$ and $P_{66} = 1$. Substituting these values in equation (29) gives

$$5\, G^{\|} \cdot R^{\|} = \left[ 2 \sum_{1}^{5} M_i N_i + 5 \sum_{1}^{6} M_6 N_6 - N_{12} \right] + \tau\, (N_{12} - N_{13}) \qquad (31)$$

The absence of $N_{16}$ term for this case is due to the fact that $P_{16} = 0$. The five planar pentagonal bases are parallel to those shown in figure (2a). Hence $\sum_{1}^{5} V_i = 0$. As argued earlier that uniqueness can be ensured by imposing $\sum_{1}^{5} N_i = -2$ to +2. Unlike the hexagonal case, the MB equivalent $\sum_1^5 N_i = 0$ or $\sum_1^5 N_i = 0$ cannot help us in reducing the form of equation (31) owing to the presence of $\tau$ or quasiperiodicity in the structure. It is clear that such bases are devoid of group-subgroup relationship with the icosahedral phase in view of severe distortion of the icosahedron. The point group of the resulting structure will be 10/m and is not a subgroup of icosahedral point group $(m\bar{3}\bar{5})$. This aspect has been dealt while discussing interfaces and twinning in quasiperiodic structures (Mandal et al. 1993; Mandal 1999, Mandal and Lele 2000).

There are experimental observations of two separate classes of decagonal phases (Pramanick et al. 2004). They are having space groups $P10_5$/mcm and P10/mmm. The presence of screw in the former case necessarily demands preservation of $\bar{5}$ symmetry in their bases. Such a choice will generate structures that are maintaining group-subgroup relationship with the icosahedral phase. Analogous to the trigonal case discussed previously, the basis vectors preserving 5-fold symmetry along the axis of distortion will help generate many 2d quasiperiodic structures. This will also include $P10_5$/mcm. Please note that for this case, the bases will satisfy the following general condition

$$\sum_1^5 V_i = 5\cos\theta_D\,|V_1|\,Z^{||} \qquad (32)$$

A choice of $\cos\theta_D = \frac{1}{2}$ has been shown to conform to the experimentally observed structures having $P\ 10_5$/mcm space group (Mandal and Lele 1991).

Owing to equation (32), the indexing would require imposition of condition on the sextuplet rather than the quintuplet (Steurrer and Kuo 1990; Steurrer et al. 1993) corresponding to $\theta_D = 90^\circ$. It is to be remembered here that different nature of group-subgroup relationships that the two classes of decagonal phases maintain with the icosahedral structures are expected to reflect upon the condition of uniqueness on indices. Such a condition for $\theta_D = 60^\circ$ can be achieved by recalling that $G^{\parallel}(22222\bar{5}) = 0$ and $R^{\parallel}(11111\bar{4}) = 0$ (Mandal and Lele 1991; Mandal et al. 2004). Hence, the conditions on indices are

$$\sum_1^6 N_i = 0 \text{ modulo } 5$$

$$= -2, -1, 0, 1, 2 \qquad (33a)$$

and $$\sum_1^6 M_i = 0 \qquad (33b)$$

For $\cos\theta_D = \frac{1}{2}$; $P_{11} = \frac{8}{15}$; $P_{12} = \frac{3\tau - 1}{15}$; $P_{12} = P_{13}$;

$P_{16} = \frac{1}{3}\sqrt{\frac{2}{5}}$; $P_{66} = \frac{1}{3}$; $(t_1/t_6) = 2\sqrt{2/5}$.

Hence, Equation (29) for this case reduces to

$$30\, G^{\parallel}\cdot R^{\parallel} = \left[16\sum_1^5 M_i N_i + 10 M_6 N_6 - 2(N_{12} - N_{13}) \right.$$

$$\left. + 8N_6 \sum_1^5 M_i + 5M_6 \sum_1^5 N_i \right] + 6\tau\,[N_{12} - N_{13}] \qquad (34)$$

Equation (34) is depiction of zone rule for $P10_5/mcm$ structure corresponding to cos $\theta_D = 1/2$. The right hand side of equation (34) is quadratic irrational. Kindly recall while arriving at zone rule in equation (30) for $\theta_D = 90^\circ$ (P10/mmm structure) the ratio of $(t_1/t_6)$ was not required. This indicates $(t_1/t_6)$ may not be fixed for this case. In contrast, quantification of zone rule for $P10_5/mcm$ types of structure requires apriori prescription of $(t_1/t_6)$ ratio. Thus the two quasiperiodic structures having periodicity in one dimension of similar type (Edagawa et al.1992; Edagawa et al.1994; Ritsch et al. 1995; Ritsch et al, 1998) demand different types of hyperlattice parameters. This distinction has to be kept in mind while discussing these structures. The two separate classes of structures display different symmetrical features. One class (5m) maintains group-subgroup relationship with the icosahedral point group whereas other (10/m) does not follow this. As a consequence of this, the path of phase transformation from the icosahedral phase for the two decagonal phases need not be same (Mandal et al. 1993; Mandal and Lele 2000).

Let us consider now the lower limit of cos $\theta = 1/\sqrt{3}$. For this choice, $P_{11} = 3/5$; $P_{12} = \frac{3}{15}\tau$; $P_{13} = -\frac{3}{15}\frac{1}{\tau}$; $P_{16} = 0$; $P_{66} = 0$. This means that $e_6$ is solely parallel to ⊥ space basis vector $W_6$. Thus the physical space or || space structures are independent of $e_6$ (or $t_6$ parameter does not play a role). In ⊥ space for this choice, cos $\phi = 0$; $|W_1| = 2/5$ and $|W_6| = 1$. In ⊥ space, pentagonal planar basis vectors will be observed. However, the transformation under five fold rotation will be non-vector like. It may be pointed out here that reverse occurs for $\theta_D = 90^\circ$ in || space. $e_6$ is solely lying in || space ($V_6$ only). This can be seen by noting cos $\phi = -\frac{1}{\sqrt{3}}$; $|W_1| = (3/5)^{1/2}$ and $|W_6| = 0$. This is the reason that $t_6$ (or hyperlattice parameter along $e_6$) is free parameter to describe the periodicities of the decagonal phases having 10-fold symmetry (Ranganathan et al. 2007). Further, periodicity of $e_6$ being parallel to $V_6$ will be equal to ~2n Å for $T_{2n}$ (n = 1,2,3 and 4) decagonal phase (Mandal and Lele 1991).

It may be recalled that for the trigonal case too, at the limit of $\cos\theta_h = \frac{1}{\sqrt{3}}$; $|A_4^{\|}| = 0$ and $|A_1^{\|}| = 1$. As a consequence of this $|A_4^{\perp}| = 1$ and $|A_1^{\perp}| = 0$. Hence the hyperlattice parameter c along $e_4$ does not play a role in || space and $e_4$ will lie in $\perp$ space.

For $\theta_h = 90^0$; $|A_1^{\|}| = \sqrt{\frac{2}{3}}$ and $|A_4^{\|}| = 1$. As a result, $|A_1^{\perp}| = 1/3$ and $|A_4^{\perp}| = 0$. Thus, $e_4$ is parallel to $A_4^{\|}$ and solely lies in this space only. For this case c of the 4d-hyperlattice cell will be equal to 3d hexagonal cell i.e. $c = c_h$. Unlike the case previously discussed for decagonal phases, no other features will be observed as $\perp$ space is one dimensional for the trigonal structures.

## 4. Novel aspects of hyper dimensional construct of trigonal, pentagonal hexagonal and decagonal phases

This presentation has discussed a unified 4d model for the description of trigonal and hexagonal phases. Such a description has facilitated the interpretation of nature of 1d aperiodic structure having trigonal symmetry. Further, hexagonal structures with 6/m point group have been shown to result from the planar basis vectors and the condition of uniqueness is shown to match with that of MB scheme. As shown in section 2, $\sum_1^3 m_i = 0$ and $\sum_1^3 n_i = 0$ leads to the transformation of 4d cell to a 3d hexagonal cell. In the limiting case of $\cos\theta_h = 1/\sqrt{3}$, the 4d cell transforms to a 3d cubic cell. Such a viewpoint is lacking in literature. This could be achieved as a general basis set of non-coplanar vectors has been taken for describing the structures of trigonal and related phases.

While discussing the structure of decagonal phases and their related phases based on 6d model (Mandal and Lele, 1989), many interesting aspects of hyperspace description have been brought out. They refer to the planar basis vectors that correspond to MB scheme of decagonal phases (Ranganathan et al. 2007) for $\theta_D = 90^\circ$.For this case, $e_6$ (or contribution of $t_6$) has been solely lying in || space. The conditions on indices $\sum_1^5 M_i = 0$ and $\sum_1^5 N_i = 0$ lead to 5d content of the 6d cell that is relevant for decagonal phases having P10/mmm symmetry. Kindly note that corresponding to above values $R_z^\perp = 0$ and $G_z^\perp = 0$ respectively in reciprocal and direct spaces of the ⊥ components. For the limiting case of distortion given by cos $\theta_D = 1/\sqrt{3}$ (analogous to the 4d counterpart), the $e_6$ (or contribution of $t_6$) is solely contained in ⊥ space. If the aforesaid conditions on indices are imposed again then $R_z^{\|} = 0$ and $G_z^{\|} = 0$. Hence 5d content of the 6d hypercell describes structures of 2d section of the decagonal phase. It is important to understand the two types of 5d content of the 6d hypercell discussed in the foregoing. For $\theta_D = 90^\circ$; $\sum_1^5 M_i = 0$; $\sum_1^5 N_i = 0$, the || space contains information for $t_1$ and $t_6$ both. In contrast, $\theta_D = \cos^{-1}(1/\sqrt{3})$; $\sum_1^5 M_i = 0$; $\sum_1^5 N_i = 0$, the ||-space has sampling from 5d hypercubic cell with parameter $t_1$ only. This will therefore have structural description in the decagonal layers only. It may be noted that a complementary description of $\theta_D = 90^\circ$ exists in ⊥ space for this case. These are being discussed to understand the limiting cases of distortion of the icosahedron along one of the six five fold axes. Table 1 summarizes essential outcomes of the 4d and 6d models to bring out the advantages of such general frame-work. The resulting structural features for different values of $\theta_h$ and $\theta_D$ are mentioned. As mentioned earlier, these subtle and novel aspects of higher dimensional modeling have not been reported in literature.

## 6. Conclusions

1. A general 4d model has been proposed for describing the structures of trigonal and their related phases on a unified basis. The Miller-Bravais scheme has been shown to result as a special case of the present framework. The limiting cases of the model [$\theta_h$ = 90° and $\theta_h = \cos^{-1}(1/\sqrt{3})$] have shown that former describes a 3d hexagonal cell as a section of 4d hypercell whereas latter leads to the transformation to a cubic cell. For the rational choice of cos $\theta_h$, periodic trigonal structure results. When this has irrational value, trigonal structures with 1d aperiodicity may be modeled.
2. Various aspects of decagonal phases having 10-fold and $10_5$ screw axis along the symmetric or periodic direction have been discussed based on 6d structure model. It has been demonstrated that they result as a consequence of distortion characterized through cos $\theta_D = 0$ and cos $\theta_D = \frac{1}{2}$ respectively. The MB equivalent conditions on indices follow if $\sum_1^5 m_i$ and $\sum_1^5 n_i$ are set to zero for $\theta_D = 90°$. The hyperlattice parameter $t_6$ (along $e_6$) seems to contribute || space only. The 3d periodicity and hyperspace periodicity along $e_6$ are the same. As a consequence of this, the decagonal phases with P10/mmm have been shown to be a 5d section of 6d hypercell. This does not seem to be the case for cos $\theta_D$ = 1/2. Thus, the two structures are different. The two extreme end of distortions corresponding to $\theta_D$ = 90° and $\cos^{-1}(1/\sqrt{3})$ have displayed complementary behaviors in || and ⊥ spaces respectively.
3. The 4d and 6d models have provided a general framework to discuss the trigonal, decagonal and their related phases respectively. Owing to the continuity of description that can be maintained through distortion parameters ($\theta_h$ and $\theta_D$), it is possible not only to understand the nature of structural phase transformations but also has the potential to interpret newer phases that may be related to parent structures.

4. Many subtle and newer aspects of hyper space description (summarized in Table 1) of various phases reported here came as an off-shoot of unified approach. This may further enhance our insight into the cut and project scheme.

**Acknowledgements**

The author is grateful to Professors N.K. Mukhopadhyay and R. Prasad (IIT, Delhi) for helpful discussion. He thanks Professors G.V.S. Sastry, R.S.Tiwary, O.N.Srivastava and S. Lele for encouragement.

Table 1. Characteristic features of trigonal and decagonal phases in terms of hyperspace description

| | **Trigonal and related phases** | **Decagonal and related phases** |
|---|---|---|
| Hyper cell | 4d orthogonal cell | 6d orthogonal cell |
| Hyper lattice parameters | a along $e_i$ (i = 1 to 3); c along $e_4$ | $t_1$ along $e_4$ (i = 1 to 5); $t_6$ along $e_6$ |
| Physical basis vectors (‖ space) | $A_1^{\parallel}$; $A_2^{\parallel}$; $A_3^{\parallel}$ and $A_4^{\parallel}$ <br> $\lvert A_1^{\parallel}\rvert = \lvert A_2^{\parallel}\rvert = \lvert A_3^{\parallel}\rvert = \sqrt{\frac{2}{3\sin^2\theta_h}}$ ; | $V_1$, $V_2$, $V_3$, $V_4$, $V_5$, $V_6$ <br> $\lvert V_1\rvert = \lvert V_2\rvert = \lvert V_3\rvert = \lvert V_4\rvert = \lvert V_5\rvert = \left(\sqrt{\frac{2}{5\sin^2\theta_D}}\right)^{1/2}$ |
| | $A_4^{\parallel} = \sqrt{\frac{1-3\cos^2\theta_h}{\sin^2\theta_h}}$ | $\lvert V_6 \rvert = \lvert V_1 \rvert \sqrt{\frac{5}{2}(1-\cos^2\theta_D)}$ |
| Pseudo basis vectors (⊥ space) | $A_1^{\perp}$ ; $A_2^{\perp}$ ; $A_3^{\perp}$ and $A_4^{\perp}$<br>$\lvert A_1^{\perp} \rvert = \lvert A_2^{\perp} \rvert = \lvert A_3^{\perp} \rvert = \sqrt{\frac{1-3\cos^2\theta_h}{3\sin^2\theta_h}}$ ;<br>$\lvert A_4^{\perp} \rvert = \sqrt{2}\cot\theta_h$ | $W_1$, $W_2$, $W_3$, $W_4$, $W_5$ and $W_6$<br>$\lvert W_1 \rvert = \lvert W_2 \rvert = \lvert W_3 \rvert = \lvert W_4 \rvert = \lvert W_5 \rvert = \sqrt{1-\lvert V_1 \rvert^2}$<br>$\lvert W_6 \rvert = \sqrt{1-\lvert V_1 \rvert^2}$ |
| Indices | Direct space : $m_i$ for i = 1 to 4<br>Reciprocal space : $n_i$ for i = 1 to 4 | Direct space : $M_i$ for i = 1 to 4<br>Reciprocal space : $N_i$ for i = 1 to 6 |
| Limiting case I | $\theta_h = 90°$; $\lvert A_1^{\parallel} \rvert^2 = 2/3$; $\lvert A_4^{\parallel} \rvert^2 = 1$; $\lvert A_1^{\perp} \rvert^2 = \frac{1}{3}$;<br>$\lvert A_4^{\perp} \rvert^2 = 0$<br>Zone rule : $\sum_1^4 n_i m_i = 0$<br>Miller-Bravais conditions<br>$\sum_1^3 m_i = 0$; $\sum_1^3 n_i = 0$<br>$c = c_h$ (3d hexagonal)<br>Frank 1965 recovered for 3d hexagonal phase | $\theta_D = 90°$; $\lvert V_1 \rvert^2 = \frac{2}{5}$; $\lvert V_6 \rvert^2 = 1$;<br>$\lvert W_1^{\perp} \rvert^2 = \frac{3}{5}$; $\lvert W_6^{\perp} \rvert^2 = 0$<br>Zone rule : $\left[2\sum_1^5 M_i N_i + 5 M_6 N_6 - N_{12}\right]$<br>$\sum_1^5 M_i = 0$; $\sum_1^5 N_i = 0$<br>$t_6$ = periodicity along 10-fold axis (Decagonal phase P10/mmm);<br><br>MB analogous of Ranganathan et al. 2007 recovered for Decagonal phase with 10 fold symmetry |
| Limiting case II | $\theta_h = \cos^{-1}\left(\frac{1}{\sqrt{3}}\right)$;<br>$\lvert A_1^{\parallel} \rvert = 1$; $\lvert A_4^{\parallel} \rvert^2 = 0$; $\lvert A_4^{\perp} \rvert = 1$ and $\lvert A_1^{\perp} \rvert = 0$ | $\theta_D = \cos^{-1}\left(\frac{1}{\sqrt{3}}\right)$;<br>$\lvert V_1 \rvert = \sqrt{3/5}$; $\lvert V_6 \rvert = 0$ |
| | 3d cubic cell in physical space for $\sum_{1}^{3} m_i = 0$; $\sum_{1}^{3} n_i = 0$. c has no role in parallel space as $e_4$ is contained in $\perp$ space | $\lvert W_1 \rvert = \sqrt{2/5}$; $\lvert W_6 \rvert = 1$<br>$t_6$ has no role in parallel space 5d cubic cell generates all features of a decagonal planar layers for $\sum_{1}^{5} M_i = 0$; $\sum_{1}^{5} N_i = 0$ |
| | $\theta_h = \cos^{-1}(1/2)$;<br>$\lvert A_1^{\parallel} \rvert = \frac{2\sqrt{2}}{3}$; $A_4^{\parallel} = \frac{1}{\sqrt{3}}$; $\lvert A_1^{\perp} \rvert = \frac{1}{3}$;<br>$\lvert A_4^{\perp} \rvert = \sqrt{\frac{2}{3}}$<br>Trigonal phase or condition for uniqueness :<br>3d cubic cell in physical space for<br>$\sum_{1}^{4} n_i = 0, \pm 1$; $\sum_{1}^{4} m_i = 0$<br>$\theta_h = \cos^{-1}\left(\frac{1}{\sqrt{5}}\right)$;<br>$\lvert A_1^{\parallel} \rvert = \sqrt{5/6}$; $\lvert A_4^{\parallel} \rvert = \frac{1}{\sqrt{2}}$; $\lvert A_1^{\perp} \rvert = \sqrt{1/6}$;<br>and $\lvert A_4^{\perp} \rvert = \frac{1}{\sqrt{2}}$<br>Trigonal phase with 1d aperiodicity | $\theta_D = \cos^{-1}(1/2)$<br>$\lvert V_1 \rvert = \sqrt{8/15}$; $\lvert V_6 \rvert = \sqrt{1/3}$<br>$\lvert W_1 \rvert = \sqrt{7/15}$; $\lvert W_6 \rvert = \sqrt{2/3}$<br>Decagonal phase $P10_5/mcm$<br>Pentagonal $\bar{5}m$<br>Condition for uniqueness :<br>$\sum_{1}^{6} N_i = 0, \pm 1, \pm 2$<br>$\sum_{1}^{6} M_i = 0$<br>$\theta_D = \cos^{-1}\left(\frac{1}{\sqrt{5}}\right)$<br>$\lvert V_1 \rvert = \lvert V_6 \rvert = \frac{1}{\sqrt{2}}$<br>Icosahedral phase recovered. |